%% file: paper.tex
\documentclass[reprint,aps,pra,twoside,superscriptaddress]{revtex4-1}

\usepackage{hyperref}

\input{fizychni_komandy.tex}

\input{notations.tex}

\usepackage{amsmath,amssymb}
\usepackage{breqn}
\usepackage{graphicx}
\usepackage{siunitx}

\makeatletter
\let\cat@comma@active\@empty
\makeatother

\usepackage{xcolor}
\definecolor{as}{rgb}{0,0,.9}
\definecolor{fwm}{rgb}{.7,0,0}

\begin{document}

\title{A superconducting detector that counts microwave photons up to two}
\author{Andrii~M.~Sokolov}
\email[E-mail: ]{andriy145@gmail.com}
\affiliation{Theoretical Physics, Saarland University, 66123 Saarbr{\"u}cken, Germany}
\affiliation{Institute of Physics of the National Academy of Sciences, pr. Nauky 46, Kyiv 03028, Ukraine}
\author{Frank K. Wilhelm}
\affiliation{Theoretical Physics, Saarland University, 66123 Saarbr{\"u}cken, Germany}

\begin{abstract}
We propose a detector of microwave photons which can distinguish the vacuum state, one-photon state, and the states with two or more photons.
Its operation is based on the two-photon transition in a biased Josephson junction and detection occurs when it switches from a superconducting to a normal state.
We model the detector theoretically.
The detector performs with more than 90\% success probability in several microseconds.
It is sensitive for the \SI{8.2}{\GHz} photons.
The working frequency could be set at the design stage in the range from about \SI 1 {\GHz} to \SI{20}{\GHz}.
\end{abstract}

\maketitle

\section{Introduction}
\label{secIntro}

Quantum optics deals with indivisible units of electromagnetic radiation on an elementary level.
It is not restricted to optical frequencies or interactions with single atoms.
In fact, the platform of circuit quantum electrodynamics based on guided microwaves and superconducting circuits containing Josephson junctions has proven successful in implementing the functionality necessary for quantum optics~\cite{you2011atomic,blais2004cavity,chen2011microwave} and to reach unparalleled coupling strengths of microwave photons to matter~\cite{forndiaz2019ultrastrong,kockum2019ultrastrong}.
It is also a successful platform for quantum computing~\cite{wilhelm2018entwicklungsstand}.
Unlike natural atoms, the matter component of circuit quantum electrodynamics could be specially tailored to perform a certain function~\cite{you2011atomic}. 
For example, one could design a counter of microwave photons which is based on Josephson junctions~\cite{chen2011microwave,opremcak2018measurement,govia2012theory,oelsner2017detection,oelsner2018switching,inomata2016single,besse2018singleshot,peropadre2011approaching,fan2014nonabsorbing}.

There are several reasons to have such a detector.
In the end of a quantum microwave experiment one usually amplifies a signal and then measures its amplitude with a homo- or a heterodyne.
To achieve a decent signal-to-noise ratio, several amplification stages are required.
Moreover, a cold stage with a quantum-limited amplifier~\cite{roy2018quantum} is used.
This requires bulky circulators and additional drive tones~(see e.g. Ref.~\cite{walter2017rapid}).
In the optical range one usually uses a photon detector, which reacts to a certain amount of energy.
Photodetectors have been also demonstrated in the microwave range~\cite{chen2011microwave,opremcak2018measurement,oelsner2017detection,oelsner2018switching,inomata2016single,besse2018singleshot}.
Josephson photon multipliers~(JPMs)~\cite{chen2011microwave,opremcak2018measurement} are especially compact and simple.
Their use allows to avoid complex and bulky amplification and promises integration with cold classical electronics~\cite{mcdermott2018quantumclassical}.
This might be useful for faster control and data acquisition, as well as for building the quantum information processing devices with more qubits.

Most designs for microwave photodetectors demonstrated so far only discriminate the vacuum state vs.~the states with a non-zero number of photons, i.e., they are called vacuum detectors~\cite{stenberg2015adaptive}.
However, for certain applications a detector that resolves the input photon number is desirable.
In the dispersive readout with a photodetector~\cite{govia2014high,opremcak2018measurement}, photon number resolution can improve fidelity in certain schemes~\cite{sokolov2016optimal}.
Other uses include optimal discrimination of coherent states~\cite{wittmann2010discrimination} and characterization~\cite{hadfield2005sfspnr} of microwave single photon sources~\cite{peng2016tuneable,forndiaz2017on}.
Detectors of microwave photons were demonstrated~\cite{johnson2010qnd} and envisioned~\cite{besse2018singleshot} that posses limited capabilities for number resolution.
However, they have a large footprint, require the use of complex pulsing sequences and are only able to distinguish a certain Fock state against all other states.

\begin{figure}
\includegraphics{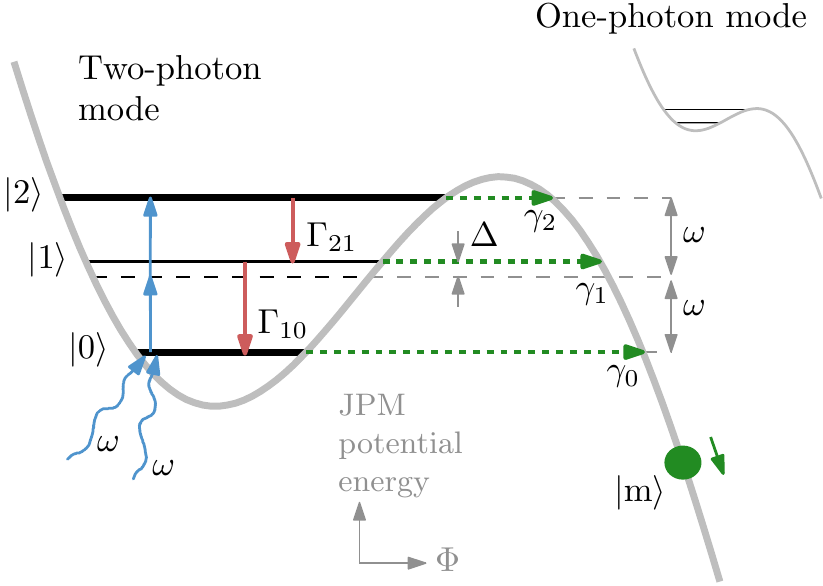}
\caption{Two modes of operation of the JPM that counts to two.
In the two-photon mode, the JPM possesses three metastable states.
A single photon rarely excites the JPM to $\ket 1$ due to detuning $\Delta$.
Two photons excite it to $\ket 2$, which then tunnels quickly to (quasi) continuum.
JPM then ``rolls'' down the potential.
This provides a macroscopic voltage on the junction, which is interpreted as a click.
In the one-photon mode, the JPM possesses two metastable states.
A single photon can deliver a click.
}
\label{figHowItWorks}
\end{figure}

We propose a photon-number resolving JPM based on the two-photon transition~(see Fig.~\ref{figHowItWorks}).
It works as follows.
First the JPM is set in the two-photon mode and its ground state is prepared.
In this mode the JPM clicks if two or more photons are present.
This can be seen as an extension of a vacuum detector.
If there are fewer than two photons, the JPM is tuned to the single-photon mode.
Here it works as a vacuum detector and fires if a photon is present.
Hence the detector discriminates the vacuum state, single-photon state, and states with two or more photons.
We present a theory of this detector in Sections~\ref{secModel}--\ref{secFastDecoherence} and evaluate its performance in Sections~\ref{secEstimates}--\ref{secCounting}.

\section{Model}
\label{secModel}

\begin{figure}
\includegraphics{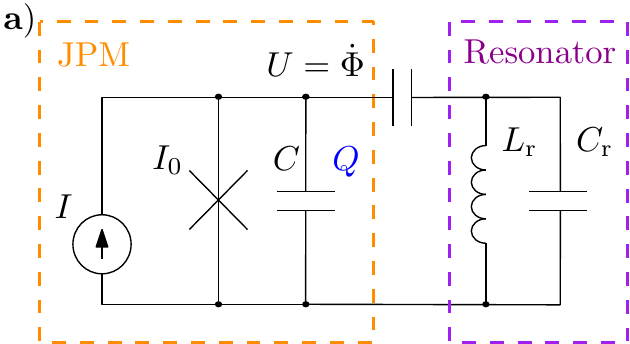}
\hfill
\includegraphics{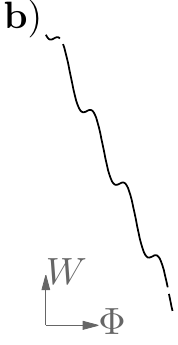}
\caption{a) Circuit diagram of a resonator mode coupled to a JPM.
The latter is a Josephson junction with a critical current $I_0$ and contact capacitance $C$.
The junction is biased with an external current $I$.
Voltage $U$ is read out by an external voltmeter.
b) Potential energy of the JPM.
}
\label{figCircuit}
\end{figure}

\begin{figure}
\includegraphics{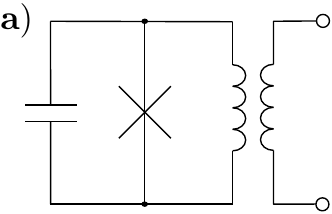}
\hskip 1.5em
\includegraphics{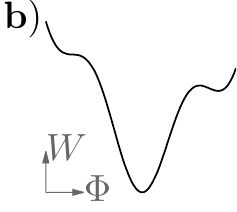}
\caption{a) Another variant of JPM schematics: a flux-biased loop with a Josephson junction.
b) Energy landscape of this JPM variant.
}
\label{figJPMCircuitLoop}
\end{figure}

In this section, we write out the Hamiltonian of our system.
Then, we treat dissipation and tunneling with the Lindblad equation formalism.
For simplicity, a current-biased Josephson junction (Fig.~\ref{figCircuit}) serves as a JPM model.
However, we also discuss why our results should be applicable for the flux-biased JPM (Fig.~\ref{figJPMCircuitLoop}).

\subsection{Hamiltonian}

We consider a resonator coupled to a JPM (see Fig.~\ref{figCircuit}).
Full system Hamiltonian is
\begin{equation}
\label{eqHamiltonian}
	H = H_\jpm + H_\coup + H_\res.
\end{equation}
Here, the resonator Hamiltonian is given by
\begin{equation}
	H_\res = \frac{Q_\res^2}{2\tilde C_\res} + \frac{\Phi_\res^2}{2L_\res},
\end{equation}
where $Q_\res$ denotes the charge on the resonator capacitance $C_\res$, and $\Phi_\res$ is the drop of quasi-flux~\footnote{Derivative of a quasiflux between circuit nodes 1 and 2 gives the respective voltage, $\dot\Phi_{12} = U_{12}$} on it~\cite{devoret1995quantum}.
A tilde denotes that a capacitance is renormalized by the JPM-resonator interaction.
The JPM Hamiltonian is of the form
\begin{equation}
\label{eqHjpm}
	H_\jpm = \frac{Q^2}{2 \tilde C} + W,
\quad
	W = -\Phi_0 I_0 \cos\frac{\Phi}{\Phi_0} - I\Phi.
\end{equation}
Here, $\Phi_0$ denotes the flux quantum.
$Q$ is the charge of the JPM capacitance $C$, $\Phi$ is its quasiflux variable.
The JPM resides in a washboard potential $W$, which is plotted in Fig.~\ref{figHowItWorks}.
The resonator and the JPM interact through a coupling capacitance $C'$.
The coupling Hamiltonian is
\begin{equation}
\label{eqCouplingHamiltonian}
	H_\coup = \frac{\tilde C'}{C C_\res} Q Q_\res.
\end{equation}
The expressions for $\tilde C$, $\tilde C_\res$, and $\tilde C'$, as well as a detailed derivation of the circuit Hamiltonian are given in Appendix~\ref{apHamiltonian}.
One can promote our canonical variables to operators.
Their commutators are:
\begin{equation}
	[\Phi, Q] = [\Phi_\res, Q_\res] = i\hbar,
\end{equation}
while the other pairs commute.
For two related circuits, a similar Hamiltonian was obtained in Ref.~\cite{anghel2019electromagnetic}, which only differs in the type of coupling.

It is convenient to rewrite the Hamiltonian~\eqref{eqHamiltonian} in terms of ladder operators.
In the Hamiltonian model, we restrict the JPM dynamics to the metastable states in a well---$\ket 0$, $\ket 1$, and $\ket 2$ in Fig.~\ref{figHowItWorks}.
For the resonator, we make a usual substitution, $\Phi_\res = \sqrt{\hbar\rho/2} \, (a + a^\dag)$ and $Q_\res = i\sqrt{\hbar/2\rho} \, (a - a^\dag)$, with $\rho = \sqrt{L_\res/\tilde C_\res}$ the renormalized impedance of the resonator.
The resulting Hamiltonian is
\begin{equation}
\label{eqHamiltonianSecQuantized}
\begin{split}
	H  = &\hbar (\omega + \Delta) \ket 1 \bra 1 
        + \hbar 2\omega \ket 2 \bra 2
        + \hbar\omega a^\dag a
\\
        &+ \hbar g_1 (\ket 1 \bra 0 a + \hc)
        + \hbar g_2 (\ket 2 \bra 1 a + \hc),
\end{split}
\end{equation}
where $g_1 = i \tilde C'(CC_\res)^{-1} \sqrt{\hbar/2\rho} \, \bra 1 Q \ket 0$ and $g_2 =  i \tilde C'(CC_\res)^{-1} \sqrt{\hbar/2\rho} \, \bra 2 Q \ket 1$.
The JPM is designed for its $0\to2$ transition frequency to match $2\omega$, where $\omega = 1/\sqrt{L_\res \tilde C_\res}$ is the resonator frequency.
The rotating-wave approximation was used in obtaining the Hamiltonian.
The coupling of the JPM to the resonator is assumed to be linear in the field quadrature, hence its matrix elements in the Fock basis couple states that are different by exactly one photon. 

\subsection{Lindbladian}

The model given so far does not take into account the interaction with the external degrees of freedom.
First, in the Hamiltonian~\eqref{eqHamiltonianSecQuantized}, we have excluded the states the system tunnels to.
Hence the tunneling is a non-unitary process in this model.
Moreover, even the non-truncated Hamiltonian~\eqref{eqHamiltonian} does not take account of the non-radiative transitions in the JPM and its dephasing.
However, it would turn out that these processes, along with the tunneling, set the JPM performance.

To model them, we use the master equation formalism~\cite{breuer2002theory}.
The Lindblad equation for our system reads
\begin{equation}
\label{eqLindbladBare}
    \dot\rho = L\rho,
\quad
	L\rho = \frac 1{i\hbar}[H, \rho] + (L_0 + L_1 + L_2)\rho.
\end{equation}
Lindbladians $L_0$, $L_1$, and $L_2$ describe the incoherent processes involving the JPM states $\ket 0$, $\ket 1$, and $\ket 2$:
\begin{align}
    L_0 &= \gamma_0 D[\ket \me \bra 0],
\\
\begin{split}
    L_i &= \Gamma_{i\,i-1} D[\ket{i-1} \bra i] + \Gamma_{i\,i} D[\ket i \bra i]
\\
        &{\phantom{=\,}} {} + \gamma_i D[\ket \me \bra i],
\quad
	i = 1, 2.
\end{split}
\end{align}
Losses in the resonator are neglected.
Here $ D[\bullet]\rho = \bullet \rho \bullet^\dag - \tfrac 1 2 [\bullet^\dag \bullet, \rho]_+$ with $[a, b]_+ = ab + ba$.
For an $i$-th excited state of the JPM, $\gamma_i$ is its tunneling rate, $\Gamma_{i\,i-1}$ is the relaxation rate, and $\Gamma_{i\,i}$ is the pure decoherence rate.
In abbreviations like these, we mean double index in a subscript.
$\ket\me$ denotes an amalgamation of the many possible states the JPM can tunnel into~\cite{govia2012theory}.
We have verified that a different tunneling model~\cite{gurvitz2011tunneling} does not change the main results of the paper.

\subsection{Flux-biased variation}

A flux-biased loop with a junction can be more convenient to operate, as it avoids voltages above the gap and hence quasiparticle production.
The circuit diagram  of this JPM variant and its energy landscape is shown in Fig.~\ref{figJPMCircuitLoop}.
Here, a click is delivered by a tunneling to bound states in the global minimum.
One aims at a regime where the global minimum resides in a wide and deep well.
Then it is unlikely for an excitation to bounce back to the local minimum and get re-emitted back to the resonator.
In fact, there is a large number of densely separated bound states, which can be treated as a continuum.
Tunneling here can be described in the same way as in the current-biased JPM we consider.
Hence, we expect same results for the flux-biased JPM variant.

\section{Effective description of the two-photon processes}

It is convenient to move to the frame where the first excited state of the JPM takes no part in the system dynamics.
The two-photon terms appear then in the Hamiltonian explicitly.
We use a Schrieffer-Wolff transform~(see Ref.~\cite{zhu2013cqedfluxonium} and references therein) to obtain the Hamiltonian in that frame.
Also, one needs to know how the relaxation processes are dressed in this picture.
Therefore, the very transform is also applied to the Lindbladian.

\subsection{Effective Hamiltonian}

One can decouple the first excited state of the JPM with the unitary transform~\cite{alexanian1995unitary}
\begin{equation}
\label{eqSchriefferWolffTransform}
    U = \exp(-\lambda_1 \ket 1 \bra 0 a + \lambda_2 \ket 2 \bra 1 a - \hc),
\end{equation}
where
\begin{equation}
\label{eqLambda12}
    \lambda_{1,2} = g_{1,2} / \Delta.
\end{equation}
Hamiltonian~\eqref{eqHamiltonianSecQuantized} is then transformed as
\begin{equation}
\label{eqEffectiveHamiltonian}
\begin{split}
    H \to U^\dag H U \approx \, &\hbar (\omega + \Delta + \chi_1) \ket 1 \bra 1
        + \hbar (2\omega - \chi_2) \ket 2 \bra 2
\\
        &{}+ \hbar \tilde g (\ket 2 \bra 0 a^2 + \hc)
\\
        &{}+ \hbar (\omega + \chi_1 \sigmaz{01} - \chi_2 \sigmaz{12}) a^\dag a
\end{split}
\end{equation}
with
\begin{gather}
    \sigmaz{ij} = \ket j \bra j - \ket i \bra i,
\quad
    \chi_i = \frac{g_i^2}{\Delta},
\\
\label{eq20Coupling}
    \tilde g = \frac{g_1 g_2}{\Delta}.
\end{gather}
By regrouping the terms in Eq.~\eqref{eqEffectiveHamiltonian}, one can check that $\chi_1$ and $\chi_2$ are the Stark shifts~\cite{blais2004cavity} per photon in the respective JPM levels.

The resulting Hamiltonian describes the system in the first order of perturbation theory.
We have neglected the terms which contribute to the $H$ matrix elements as $\lambda_{1,2}^2 \nch$ or $\lambda_1 \lambda_2 \nch$, where $\nch$ is a characteristic number of photons in the resonator.
Hence the Hamiltonian~\eqref{eqEffectiveHamiltonian} holds if
\begin{equation}
\label{eqValidityOfEffH}
	\lambda_{1,2}^2 \nch \ll 1.
\end{equation}

A transform is known~\cite{wu1996effective}, that \emph{exactly} decouples the first excited state of a three-level atom interacting with a resonator mode.
However, it does not accomplish this in the presence of environment and is hence not useful here.

\subsection{Interaction picture}

It is convenient to move to the interaction picture with a unitary transform $U_\inter = \exp -iH_0 t / \hbar$, where $H_0$ is the Hamiltonian of the qubit and the resonator including the parametric interaction terms.
This gives rise to
\begin{gather}
\label{eqHamiltonianInteractionPic}
	H \to U_\inter^\dag H U_\inter = \hbar \tilde g \ket 2 \bra 0 e^{irt} a^2 + \hc,
\\
\label{eqDefR}
	r = \chi_1(N - 2 \sigmaz[01]) - \chi_2 (1 + N - 2\sigmaz[12]),
\quad
	N = a^\dag a.
\end{gather}
It was used that $\ket 2 \bra 0\, \to \ket 2 \bra 0\, \exp i (2\omega - \chi_2(1+N) + \chi_1 N) t$ and $a^2 \to a^2 \exp -2i(\omega + \chi_1 \sigmaz{01} - \chi_2\sigmaz{12})t$.

In the interaction picture, the non-diagonal elements of the density matrix (coherences) do not oscillate with a high frequency.
This simplifies the differential equations that govern the matrix elements.
What is more important, in the interaction picture decoherence becomes the fastest process.
This would allow us to make crucial approximations in Sec.~\ref{secFastDecoherence}.

Before that, one needs to check how the Lindbladian changes with the transition to the working frame by the unitary transform $U$, which is given in Eq.~\eqref{eqSchriefferWolffTransform}.

\subsection{Effective Lindbladian}

Transition to another frame with the unitary transform $U$ changes the rates of non-unitary processes.
In that frame, a resonator photon gets dressed by the JPM, thus acquiring new channels of tunneling and decay.
One needs to find the Lindbladian in our working frame.

While the density matrix transforms by $\rho \to U \rho U^\dag$, elements of Lindbladians transform as
\begin{equation}
    \ket i \bra j \to U^\dag \ket i \bra j U,
\quad
    i, j = 0,1,2,\me.
\end{equation}
An explicit form of the transformed Lindbladian is given in Appendix~\ref{apLindblad}.

\section{Equations for the click probability}

Probability of the detector click is given by the occupation of $\ket\me$ disregarding the resonator state,
\begin{equation}
\label{eqPGeneral}
    P = \sum_{N=0}^\infty \rho_{N\me, N\me}.
\end{equation}
In this section, we write out the exact equations that allow to calculate $P$.

First the equation on $\rho_{\me\me}$ is given.
Here and in what follows, we use the abbreviations
\begin{equation}
\label{eqMatrixElementsDeciphered}
    \rho_{M i, N j} = \bra M \rho_{ij} \ket N,
\quad
	\rho_{ij} = \bra i \rho \ket j,
\end{equation}
where $i,j = 0,1,2,\me$ index the JPM states while $M$ and $N$ index the Fock states of the resonator.
Projecting the dressed Lindbladian~\eqref{eqLindbladianDressed} on $\ket\me$ gives rise to
\begin{equation}
\label{eqRhomm}
\begin{split}
    \dot\rho_{\me\me}
		= &\gamma_0 \rho_{00} + \gamma_1 \rho_{11} + \gamma_2 \rho_{22}
\\
		 &{} + (\gamma_1 \lambda_2 \rho_{12} a - \gamma_0 \lambda_1 \rho_{01} a
\\
		 & \phantom{{}+(} {}-
				\gamma_2 \lambda_2 a \rho_{12} + \gamma_1 \lambda_1 a \rho_{01}
			 + \hc).
\end{split}
\end{equation}
The equation is given up to and including terms of order $\lambda_1$ and $\lambda_2$.

To complete it, one needs equations on $\dot\rho_{00}$, $\dot\rho_{11}$, and $\dot\rho_{22}$ with the same accuracy.
For $\dot\rho_{12}$ and $\dot\rho_{01}$ zeroth approximation in $\lambda_1$ and $\lambda_2$ would suffice.
It is convenient to use the reduced $r$~\eqref{eqDefR},
\begin{equation}
	r_0 = \bra 0 r \ket 0,
\end{equation}
that considers the JPM in the ground state $\ket 0$ and acts solely on the resonator.
Equations~\eqref{eqLindbladianDressed}~and~\eqref{eqHamiltonianInteractionPic}--\eqref{eqDefR} yield
\begin{align}
\label{eqRho00}
\nonumber
	\dot\rho_{00} &= (i\tilde g \rho_{02} e^{ir_0 t} a^2 + \hc)
					- \gamma_0\rho_{00} + \Gamma_{10}\rho_{11}
\\
					& \phantom{{}=} {}+ \bra 0 L^{(1)} \ket 0,
\\
	\dot\rho_{11} &= -(\gamma_1 + \Gamma_{10})\rho_{11} + \Gamma_{21}\rho_{22}
					+ \bra 1 L^{(1)} \ket 1,
\\
\label{eqRho22}
\nonumber
	\dot\rho_{22} &= (-i\tilde g e^{ir_0t} a^2 \rho_{02} + \hc)
					- (\gamma_2 + \Gamma_{21})\rho_{22}
\\
					& \phantom{{}=} {}+ \bra 2 L^{(1)} \ket 2
\end{align}
in the first order in $\lambda_1$ and $\lambda_2$.
Next we express
\begin{align}
\label{eqRho01}
	\dot\rho_{01} &= -i\tilde g a^{\dag2} e^{-ir_0t} \rho_{21}
					-\frac 1 2 d_{01} \rho_{01} + O(\lambda_1 + \lambda_2),
\\
\label{eqRho12}
	\dot\rho_{12} &= i\tilde g \rho_{10} a^{\dag2} e^{-ir_0t}
					- \frac 1 2 d_{12} \rho_{12} + O(\lambda_1 + \lambda_2)
\end{align}
in terms of the $\rho_{00}$, $\rho_{11}$, and $\rho_{22}$, as well as
\begin{multline}
\label{eqRho02}
	\dot\rho_{02} = i\tilde g \rho_{00} a^{\dag2} e^{-ir_0t}
					- i\tilde g a^{\dag2} e^{-ir_0t} \rho_{22}
\\
					- \frac 1 2 d_{02} \rho_{02}
					+ \bra 0 L^{(1)} \ket 2
					+ O(\lambda_1^2 + \lambda_2^2 + \lambda_1\lambda_2),
\end{multline}
where
\begin{align}
\label{eqFullDecoherence01}
	d_{01} &= \gamma_0 + \gamma_1 + \Gamma_{10} + \Gamma_{11},
\\
\label{eqFullDecoherence12}
	d_{12} &= \gamma_1 + \gamma_2 + \Gamma_{10} + \Gamma_{21} + \Gamma_{22},
\\
\label{eqFullDecoherence02}
	d_{02} &= \gamma_0 + \gamma_2 + \Gamma_{21} + \Gamma_{22}
\end{align}
are the full decoherence rates of the $0\to1$, the $1\to2$, and the $0\to2$ transitions, respectively.
Due to the form of Eqs.~\eqref{eqRho00}~and~\eqref{eqRho22}, we have calculated  $\dot\rho_{02}$ in the first order in $\lambda_1$ and $\lambda_2$.

It is not hard to write out a full set of equations to calculate $\rho_{\me\me}$ and $P$.
To do that, one uses the expressions for $L^{(1)}$ matrix elements from Appendix~\ref{apLindblad} and projects Eqs.~\eqref{eqRhomm} and~\eqref{eqRho00}--\eqref{eqRho02} on the photon number states.
However, in the regime the device operates well, much simpler equations can be used.

\section{Fast decoherence}
\label{secFastDecoherence}

If there are two photons in the resonator, the JPM should fire as fast as possible.
After the photons excite the JPM, it should tunnel immediately.
More precisely, this should happen much faster than the excitation bounces back coherently to the cavity or the JPM relaxes non-radiatively.
In this regime, the JPM decoheres instantaneously;
hence the system state is determined by the probabilities of the excitation to occupy either the cavity or the JPM.
Here we obtain the rate equations for the case of fast decoherence.

For that case we assume that
\begin{equation}
\label{eqFastDecoherence}
	\tilde\Gamma_1 + \Gamma_{11} \gg t^{-1},
\quad
	\tilde\Gamma_2 + \Gamma_{22} \gg \tilde\Gamma_1, t^{-1},
\end{equation}
where $t$ is the time we observe the system and
\begin{equation}
	\tilde\Gamma_1 = \gamma_1 + \Gamma_{10},
\quad
	\tilde\Gamma_2 = \gamma_2 + \Gamma_{21}.
\end{equation}
By Eqs.~\eqref{eqFastDecoherence}, and given that
\begin{equation}
\label{eqTunnelingRatesHierarchy}
	\gamma_0 \ll \gamma_1 \ll \gamma_2,
\end{equation}
Eqs.~\eqref{eqFullDecoherence01} and~\eqref{eqFullDecoherence12} yield $d_{01} \approx \tilde\Gamma_1 + \Gamma_{11}$ and $d_{12} \approx \tilde\Gamma_2 + \Gamma_{22}$.
Moreover, at time $t$ coherences have already died out,
\begin{equation}
\label{eqVanishingCoherences}
	\rho_{01} \approx \rho_{12} \approx 0, 
\end{equation}
which follows from the form of Eqs.~\eqref{eqRho01}--\eqref{eqRho12} and the conditions~\eqref{eqFastDecoherence}.
Equation~\eqref{eqRhomm} then simplifies to
\begin{equation}
\label{eqRhommFastDecoherence}
    \dot\rho_{\me\me}
		= \gamma_0 \rho_{00} + \gamma_1 \rho_{11} + \gamma_2 \rho_{22}.
\end{equation}

One can show the system is then governed by rate equations.
First we express $\rho_{02}$ in terms of the probabilities $\rho_{00}$ and $\rho_{22}$.
The formal solution of Eq.~\eqref{eqRho02} reads
\begin{multline}
	\rho_{02}(t)
		\approx \rho_{02}(0)e^{-\frac 1 2 (\tilde\Gamma_2 + \Gamma_{22})t}
		+ i\tilde g \int_0^t dt'
			e^{-\frac 1 2 (\tilde\Gamma_2 + \Gamma_{22})(t-t')}
\\
		\times [\rho_{00} a^{\dag2} e^{-ir_0 t'}
				- a^{\dag2} e^{-ir_0 t'} \rho_{22} e^{\tilde\Gamma_2t'}
														 e^{-\tilde\Gamma_2t'}].
\end{multline}
The first term in the right-hand side vanishes due to Eqs.~\eqref{eqFastDecoherence}.
We assume $a^\dag(t')$, $e^{-ir_0 t'}$, $\rho_{00}(t')$, and $\rho_{22}(t') e^{\tilde\Gamma_2t'}$ to change slowly in comparison to the rate $\tilde\Gamma_2$.
Taking them out of the integral allows one to perform the integration, which yields
\begin{equation}
	\rho_{02}(t) \approx i \frac{2\tilde g}
								{\tilde\Gamma_2 + \Gamma_{22}}
						\rho_{00} a^{\dag2} e^{-ir_0 t}.
\end{equation}
Substituting this into Eqs.~\eqref{eqRho00}--\eqref{eqRho22} and projecting them on the resonator Fock states gives rise to
\begin{equation}
\begin{split}
   \label{eqRhoDiffEqs}
   \dot\rho_{N\,0, N\,0} &= { - B_{N\,N-2} \rho_{N\,0, N\,0}}
                            - \gamma_0 \rho_{N\,0, N\,0}
\\
                            &\phantom{{}=} {}+ \Gamma_{10} \rho_{N\,1, N\,1},
   \quad N \geq 2
\\
   \dot\rho_{N-2\,2, N-2\,2} &= {B_{N\,N-2} \rho_{N\,0, N\,0}}
\\
		&\phantom{{}=} {}- \tilde\Gamma_2 \rho_{N-2\,2, N-2\,2},
   \quad N \geq 2
\\
   \dot\rho_{N\,0, N\,0} &= -\gamma_0 \rho_{N\,0, N\,0}
                            + \Gamma_{10}\rho_{N\,1, N\,1},
   \quad N = 0, 1
\\
   \dot\rho_{N\,1, N\,1} &= -\tilde\Gamma_1\rho_{N\,1, N\,1}
                            + \Gamma_{21}\rho_{N\,2, N\,2}.
\end{split}
\end{equation}
It was used that the matrix elements $\bra i L^{(1)} \ket i \approx 0$ for $i = 0,1,2$ due to Eq.~\eqref{eqVanishingCoherences}.
We have defined
\begin{equation}
\label{eqNNminus2Absorption}
	B_{N\,N-2} = \frac{4\tilde g^2}{\tilde\Gamma_2 + \Gamma_{22}} N(N-1)
\end{equation}
the ratio of absorption of two photons from an $N$-photon state.
$\Gamma_{N\,N-2}$ is also the stimulated emission rate;
however, Eqs.~\eqref{eqRhoDiffEqs} do not contain stimulated emission terms, as the stimulated emission is slow compared to the competing processes.
One can figure out from Eqs.~\eqref{eqRhoDiffEqs} that the condition
\begin{equation}
\label{eqSlowStimulatedEmission}
	B_{N\,N-2} \ll \tilde\Gamma_2 + \Gamma_{22}
\end{equation}
should hold, as we have assumed $\rho_{00}$ and $\rho_{22}$ to change slowly.
Also, as $e^{-ir_0 t'}$ is assumed to change slowly as well, the condition
\begin{equation}
\label{eqSmallStarkShift}
	\chi_2 N_{\text{max}} \ll \tilde\Gamma_2 + \Gamma_{22}
\end{equation}
should hold.
It was taken into account that $\chi_2 - \chi_1 \sim \chi_2$ as
\begin{equation}
\label{eqEstimateRatiog2g1}
	g_2 \approx \sqrt 2 g_1
\end{equation}
in the harmonic approximation of the JPM potential.
$N_{\text{max}}$ in Eq.~\eqref{eqSmallStarkShift} is the highest number of a Fock state such that its occupation is not negligible.
The condition~\eqref{eqSmallStarkShift} is easy to interpret in the laboratory frame.
It makes sure that the two photon transition is not detuned from the Stark-shifted second excited level more than by its linewidth.
This interpretation suggests that the condition might be weakened to use the ``$<$'' inequality sign.

As manifested by Eqs.~\eqref{eqRhoDiffEqs}, dressing does not change the rates of non-unitary processes if the decoherence is fast;
this can be explained as follows.
Consider $\ket{1\, 0}$ the dressed state of a photon and the ground-state JPM.
In terms of the bare states, it is a photon entangled with the excited JPM, $\ket{1\, 0} \approx \ket{1\, 0}_\bare  + \lambda_1 \ket{0\, 1}_\bare$.
Admixture of the bare excited JPM adds its decay channels to the dressed state.
However, due to the rapid decoherence, the state collapses to a statistical mixture.
The addition to the decay rate is then of order of $\lambda_1^2$, which is negligible.

In the next subsections we calculate the click probabilities for vacuum, single-photon, and two-photon inputs.
In the two-photon mode, a click should be delivered if more than one photon dwells in the resonator;
no click should occur in the opposite case.
Clicks that do occur for vacuum or single-photon inputs we call false counts.

\subsection{Vacuum input}

Here we determine the probability of a JPM click in the case there are no photons in the resonator.

First we determine the initial state of the system.
In the laboratory frame, both the JPM and the cavity are in the ground state at the initial instant:
\begin{equation}
	\ket{\Psi_\bare(0)} = \ket{0\, 0}.
\end{equation}
So are they in our working frame,
\begin{equation}
	\ket{\Psi(0)} = U^\dag \ket{\Psi_\bare(0)} = \ket{0\, 0},
\end{equation}
where $U$ is defined in Eq.~\eqref{eqSchriefferWolffTransform}.
Therefore,
\begin{equation}
\label{eqVacuumInitialRho}
	\rho(0) = \ket{0\,0} \bra{0\, 0}.
\end{equation}

For this case, Eqs.~\eqref{eqRhoDiffEqs},~\eqref{eqPGeneral} and~\eqref{eqRhommFastDecoherence} simplify to
\begin{equation}
	\dot \Pfalse = \gamma_0 \rho_{00,00}
\quad \text{and} \quad
	\dot\rho_{00,00} = -\gamma_0 \rho_{00,00}.
\end{equation}
With the initial conditions given by Eq.~\eqref{eqVacuumInitialRho} and
\begin{equation}
\label{eqInitialP}
	\Pfalse(0) = 0,
\end{equation}
these equations yield
\begin{equation}
\label{eqP0}
	\Pfalse(t) = 1 - e^{-\gamma_0 t}
\end{equation}
with $\gamma_0$ the false count rate.
The JPM can tunnel even while in the ground state, hence delivering a false count.

\subsection{One-photon input}

Analogously to the previous case, one can determine the initial state.
In the laboratory frame
\begin{equation}
	\ket{\Psi_\bare(0)} = \ket{1\, 0},
\end{equation}
while in the working frame
\begin{gather}
\begin{aligned}
	\ket{\Psi(0)} = &U^\dag \ket{\Psi_\bare(0)}
\\
\label{eq1PhotonInitialPsi}
		= &\ket{1\, 0} + \lambda_1 \ket{0\, 1}
		+ O(\lambda_1^2) + O(\lambda_2^2),
\end{aligned}
\\
\begin{split}
\label{eq1PhotonInitialRhoFull}
	\rho(0) = \ket{1\,0} \bra{1\, 0} + \lambda_1 \ket{1\,0} \bra{0\, 1}
			+ \lambda_1 \ket{0\,1} \bra{1\, 0}
\\
			+ O(\lambda_1^2) + O(\lambda_2^2).
\end{split}
\end{gather}
Recall that, in our working frame, there is no interaction with the JPM first excited state.
However, in this frame, a bare photon acquires a part of it according to Eq.~\eqref{eq1PhotonInitialPsi}.
This may cause a click if the excitation from the first level tunnels.

In the limit of fast decoherence, the dressed initial state coincides with the bare one.
Due to Eq.~\eqref{eqVanishingCoherences}, coherences vanish on times~\eqref{eqFastDecoherence} we are interested in and
\begin{equation}
\label{eq1PhotonInitialRho}
	\rho(0) \approx \ket{1\,0} \bra{1\, 0}.
\end{equation}

Solving Eqs.~\eqref{eqRhoDiffEqs},~\eqref{eqPGeneral} and~\eqref{eqRhommFastDecoherence} with the initial conditions given by Eqs.~\eqref{eq1PhotonInitialRho} and~\eqref{eqInitialP} yields
\begin{equation}
\label{eqP1}
	\Pfalse(t) = 1 - e^{-\gamma_0 t}.
\end{equation}
The false count rate for the single-photon input is the same as for the vacuum input.
This could be explained as follows.
As commented before, the one-photon admixture in Eq.~\eqref{eq1PhotonInitialRhoFull} may deliver a click.
However, it relies on the system coherence.
The coherence dies out momentarily and the admixture decays before the JPM excitation can tunnel.

This does not hold in the next order of perturbation theory.
Luckily, there is a simple way to estimate the next-order false count rate.
This rate will provide the limit of applicability of Eq.~\eqref{eqP1}.

\subsubsection{Limit on the measurement time \mbox{due to the excitation of the first excited state}}

Let us calculate the tunneling rate due to the one photon transition in the next order of perturbation theory.
As the conditions~\eqref{eqFastDecoherence} of the fast decoherence are secured, one can argue in terms of probabilities and transition rates.
From Eq.~\eqref{eq1PhotonInitialPsi}, probability of the JPM residing in the first excited state is
\begin{equation}
\label{eqProbabilityOneExcitationInJPM}
	\bra 1 \rho(0) \ket 1
		= \lambda_1^2 + O(\lambda_1^3) + O(\lambda_2^3).
\end{equation}
According to Eqs.~\eqref{eqRhoDiffEqs}, the first excited state tunnels with rate $\gamma_1$.
Therefore, for the initial state $\rho(0)$, the rate of the first-level tunneling is $\lambda_1^2 \gamma_1$.
Note the Eqs.~\eqref{eqRhoDiffEqs} are obtained up and including terms of order of $\lambda_{1,2}$ only.
However, higher-order terms in the equations only give rise to corrections of order beyond $\lambda_{1,2}^2$ in the rate.

The tunneling rate via the first excited state sets the limit of validity of Eq.~\eqref{eqP1},
\begin{equation}
\label{eqValidityTimeFirstOrderPerturbation}
	t \ll \gamma_1^{-1} \lambda_1^{-2}.
\end{equation}
For the two photon input we consider below, the limit of validity is the same.
It can be obtained analogously.

\subsection{Two-photon input}

To determine initial conditions, one applies the same reasoning as for the vacuum and the one-photon input.
This gives rise to
\begin{equation}
\label{eqRhoTwoPhoton}
	\rho(0) = \ket{2\,0} \bra{2\, 0} + \lambda_1 \ket{2\,0} \bra{1\, 1}
			 + \lambda_1 \ket{1\,1} \bra{2\, 0}.
\end{equation}
Due to the fast decoherence, the initial state should be approximated as
\begin{equation}
\label{eq2PhotonInitialRho}
	\rho(0) \approx \ket{2\,0} \bra{2\, 0}.
\end{equation}
This can be shown analogously to the case of one-photon input.

\begin{figure}
\includegraphics{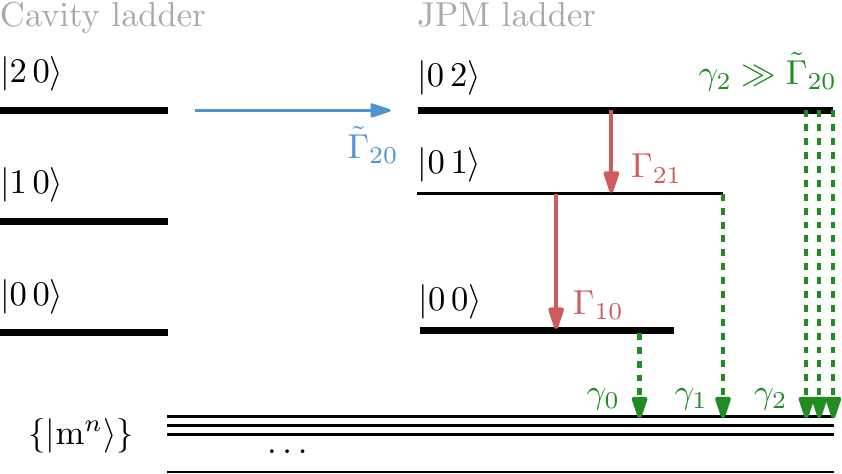}
\caption{Two-photon absorption in the limit of fast decoherence and tunneling.
$\{\ket{\me^n}\}$ are the states the JPM can tunnel to.}
\label{figRates}
\end{figure}

Equations~\eqref{eqRhoDiffEqs},~\eqref{eqPGeneral} and~\eqref{eqRhommFastDecoherence} become
\begin{align}
\label{eqRho2020}
	\dot\rho_{20,20} &= -B_{20} \rho_{20,20}
						- \gamma_0 \rho_{20,20},
\\
\label{eqRho0202}
	\dot\rho_{02,02} &= B_{20} \rho_{20,20}
						- \tilde\Gamma_2\rho_{02,02},
\\
	\dot\rho_{01,01} &= -\tilde\Gamma_1\rho_{01,01}
						+ \Gamma_{21} \rho_{02,02},
\\
\label{eqRho0000}
	\dot\rho_{00,00} &= -\gamma_0\rho_{00,00} + \Gamma_{10}\rho_{01,01},
\\
\label{eqP2Diff}
	\dot \Pbright &= \gamma_0\rho_{20,20} + \gamma_2\rho_{02,02}
		+ \gamma_1\rho_{01,01} + \gamma_0\rho_{00,00}
\end{align}
with
\begin{equation}
\label{eq20Absorption}
	B_{20} = 8\tilde g^2 /(\tilde\Gamma_2 + \Gamma_{22})
\end{equation}
the two-photon absorption rate.
The rate equations are illustrated in Fig.~\ref{figRates}.
Equations, similar to Eqs.~\eqref{eqRho2020}--\eqref{eqP2Diff} were obtained in Ref.~\cite{govia2012theory} for the one-photon transition in a two-state JPM well.
As compared to the reference, our equations lack the stimulated emission terms.
This is explained for Eqs.~\eqref{eqRhoDiffEqs}.
Moreover, the ground level tunneling was not accounted for in the reference.

We solve Eqs.~\eqref{eqRho2020}--\eqref{eqP2Diff} by carrying out the Laplace transform.
The initial conditions are given by Eqs.~\eqref{eq2PhotonInitialRho} and $\Pbright(0) = 0$.
The solution in the Laplace domain is
\begin{multline}
	\tilde \Pbright(s) = \frac{\gamma_0 B_{20} \Gamma_{21} \Gamma_{10}}
						{s(s+\gamma_0)(s+\tilde\Gamma_1)\Delta_2}
				+ \frac{\gamma_1 B_{20} \Gamma_{21}}
						{s(s+\tilde\Gamma_1)\Delta_2}
\\
				+ \frac{\gamma_0(B_{20} + \tilde\Gamma_2) 
						+ \gamma_2 B_{20}}
						{s\Delta_2}
				+ \frac{\gamma_0}{\Delta_2},
\end{multline}
where
\begin{equation}
	\Delta_2 = (s+\tilde\Gamma_2)(s+B_{20} + \gamma_0).
\end{equation}

Now we find an expression for the click probability in the time domain.
It is found by calculating the inverse Laplace transform,
\begin{equation}
	\Pbright(t) = \frac 1 {2\pi i}
		\int_{\sigma-i\infty}^{\sigma+i\infty} ds\, e^{st} \tilde \Pbright(s).
\end{equation}
By carrying out the integrals and doing some approximations, one arrives at
\begin{multline}
\label{eqP}
	\Pbright(t) = 1 - e^{-B_{20}t}
			 - \frac{\Gamma_{21}}{\gamma_2 + \Gamma_{21}}
				\frac{\Gamma_{10}}{\gamma_1 + \Gamma_{10}} e^{-\gamma_0 t}.
\end{multline}
We used the condition~\eqref{eqTunnelingRatesHierarchy} and
\begin{equation}
\label{eqTunnelingFrom20Negligible}
	\gamma_0 \ll B_{20} \ll \tilde\Gamma_2,
\end{equation}
where the last inequality is a more stringent version of the condition~\eqref{eqSlowStimulatedEmission}.
This allowed to drop the terms proportional to $B_{20}/\tilde\Gamma_2$ and $\gamma_{0,1}/\tilde\Gamma_2$.
These terms are negligibly small in comparison to the second term in the equation.
While it would turn out the last term is also small, it decays much slower than the second one.
Hence it is considerable for longer times $t$.
Equation~\eqref{eqP} holds for the times~\eqref{eqFastDecoherence} coherence has already vanished.

One can interpret Eq.~\eqref{eqP}.
The second term there is the population of the state $\ket{2\,0}$ of the resonator in the two-photon Fock state and the JPM in the ground state.
Tunneling from this state is negligible due to Eq.~\eqref{eqTunnelingFrom20Negligible}.
After an excitation transfers from $\ket{2\, 0}$ to $\ket{0\, 2}$ with the rate $B_{20}$, it tunnels immediately due to the condition~\eqref{eqSlowStimulatedEmission}.
Hence $1 - \exp(-B_{20}t)$ is the tunneling probability for the times before the resonator is depleted.
Afterwards, the third term in Eq.~\eqref{eqP} starts to matter.
While absorbing photons, the JPM can also relax to its first excited state $\ket 1$.
After all photons are absorbed, the JPM relaxes to $\ket 1$ with a small probability $\Gamma_{21}/(\gamma_2 + \Gamma_{21})$.
From $\ket 1$ the JPM relaxes to the ground state with the probability $\Gamma_{10}/(\gamma_1 + \Gamma_{10})$.
There it is stuck due to the slow ground-state tunneling of rate $\gamma_0$, which only becomes substantial for the longer times.
While a tunneling can also occur from $\ket 1$, this mostly happens while the resonator is not yet depleted and the tunneling from $\ket{0\, 2}$ is ongoing.
Due to the condition~\eqref{eqTunnelingRatesHierarchy}, this process is much faster than the tunneling from $\ket 1$ and the respective term does not play a role in Eq.~\eqref{eqP}.

\subsection{Error probability}
\label{secError}

One can now calculate the probability of false discrimination between the state with $N=2$ photons and the states with $N=1$ or $N=0$ photons.
This error is expressed as
\begin{equation}
	\varepsilon = P_{0,1} \Pfalse + P_2 (1-\Pbright)
\end{equation}
where $P_N$ is a probability of an input state with $N$ photons to occur.
$\Pbright$ denotes a probability of a bright count---i.e., a probability of registering a two photon state when it dwells in the resonator.
It was taken into account that the probability of a false count $\Pfalse$ is the same for both $N=0$ and $N=1$.

If we know nothing about the resonator state beforehand, $P_{0,1} = P_2 = 1/2$.
Using the expressions~\eqref{eqP0} and~\eqref{eqP1} for $\Pfalse$ and Eq.~\eqref{eqP} for $\Pbright$ yields
\begin{equation}
\label{eqError}
	\varepsilon = \frac 1 2 (1 + e^{-B_{20} t}
							  + \Big(\frac{\Gamma_{21}}{\tilde\Gamma_2}
								 \frac{\Gamma_{10}}{\tilde\Gamma_1} - 1\Big)
															e^{-\gamma_0 t}).
\end{equation}
The error probability is plot in Fig.~\ref{figError}.

With Eq.~\eqref{eqError}, it is possible to find the minimal error and the optimal waiting time $t$.
At
\begin{equation}
\label{eqOptimalTime}
	t \approx \frac 1 {B_{20}} \ln\frac{B_{20}}{\gamma_0}
\end{equation}
one attains the minimal error
\begin{equation}
\label{eqMinimalError}
	\varepsilon \approx \frac{\gamma_0}{2B_{20}}
							\Big(1 + \ln\frac{B_{20}}{\gamma_0}\Big)
		+ \frac 1 2 \frac{\Gamma_{21}}{\tilde\Gamma_2}
					\frac{\Gamma_{10}}{\tilde\Gamma_1}.
\end{equation}
One can check the expression is the same if the condition with $\gamma_0$ in Eq.~\eqref{eqTunnelingFrom20Negligible} is not used in obtaining Eq.~\eqref{eqP}.

\subsection{More than two photons in the input}

For the case there are $N>2$ photons in the cavity, a two-photon transition occurs, leaving $N-2$ photons in the cavity.
To describe this, one only need to change the state labels and $B_{20} \to B_{N\,N-2}$ in Eqs.~\eqref{eqRhoTwoPhoton}--\eqref{eqMinimalError}.
The bright count probability $\Pbright$ improves, as $B_{N\,N-2} > B_{20}$ by Eq.~\eqref{eqNNminus2Absorption}.
By the same reason, the error $\varepsilon$ gets smaller if one needs to discriminate a state with $N>2$ photons against the states with one or no photons.
Moreover, the error is smaller even if $N$ breaks the condition~\eqref{eqValidityOfEffH} but the requirement~\eqref{eqSmallStarkShift} still holds.
In that case, additional clicks are provided by the single-photon transition and the subsequent tunneling from the first level.

\section{Distinguishing a multi-photon state}
\label{secEstimates}

In this section, example parameters for the JPM in the two-photon mode are provided.
For those parameters, we estimate the probabilities of bright and false counts, the time to distinguish a multi-photon state, and the probability of false discrimination.

First let us summarize the requirements for our JPM to work as described above.
The energy of the junction plasma oscillations should much exceed that of a thermal excitation, $\hbar\omega_\plasma \gg k_{\text B} T$, where $T$ is the temperature of the JPM environment.
On the other hand, we do not want to spur quasiparticles while exciting the JPM.
Hence
\begin{equation}
\label{eqBigGap}
	\omega_\plasma \ll \Delta_{\text{gap}},
\end{equation}
where $\Delta_{\text{gap}}$ is the superconductor gap.
Furthermore, the effective Lindbladian~\eqref{eqLindbladianDressed} we have used is correct if the conditions~\eqref{eqValidityOfEffH} and~\eqref{eqValidityTimeFirstOrderPerturbation} hold.
Finally, we have required the JPM to decohere fast by the conditions~\eqref{eqFastDecoherence} and~\eqref{eqSlowStimulatedEmission}--\eqref{eqSmallStarkShift}.

\newcommand\T{\rule{0pt}{2.6ex}}       
\newcommand\B{\rule[-1.2ex]{0pt}{0pt}} 

\begin{table*}
\caption{Parameters and estimates for the JPM, as well as its performance in the detection of the two-photon state.
The junction parameters are from Ref.~\cite{martinis2005dielectric}.
Coupling strength $g_1$ is chosen as described in the text.
The bright $\Pbright$~\eqref{eqP} and the false $\Pfalse$ count probabilities~[Eqs.~\eqref{eqP0} and~\eqref{eqP1}] are given for the optimal waiting time $t$~\eqref{eqOptimalTime}.
}

\begin{center}
\label{tablParameters}
\begin{ruledtabular}
\begin{tabular}{ l c c c c c c c c c c c c c c r }
\multicolumn 6c{Parameters} & \multicolumn 7c{Estimated values}
	& \multicolumn 3c{Performance}
\B
\\
\cline{1-6} \cline{7-13} \cline{14-16}
\T
$C$ & $I_0$ & $I/I_0$ & $\Gamma_{10}/2\pi$ & $\Gamma_{22}/2\pi$ & $g_1/2\pi$
	& $\gamma_0/2\pi$ & $\gamma_1/2\pi$ & $\gamma_2/2\pi$
	& $\omega/2\pi$ & $\Delta/2\pi$ & $B_{20}/2\pi$
	& $N_{\text{max}}$
	& t & $\Pfalse$ & $\Pbright$
\\
 (pF) & ($\mu$A) &  & (kHz) & (MHz) & (MHz)
	& (Hz) & (kHz) & (MHz)
	& (GHz) & (MHz) & (MHz)
	&
	& ($\mu$s) & \% & \%
\B
\\
\cline{1-6} \cline{7-13} \cline{14-16}
\T
 2 & 10 & 0.97987 & 318 & 2.1 & 18.9
	& 37 & 54 & 41
	& 8.2 & 194 & 0.35
	& 14
	& 4.2 & 0.1 & 98.6
\end{tabular}
\end{ruledtabular}		
\end{center}

\end{table*}

It is convenient to introduce the Josephson energy
\begin{equation}
\label{eqWj}
	\Wj = \frac{I_0\Phi_0}{2\pi}
\end{equation}
and the capacitive energy
\begin{equation}
	\Wc = \frac{e^2}{2C}.
\end{equation}

We want to fit three levels in the well.
Besides, the third level is kept quite far from the top of the barrier.
This would make our estimates for the tunneling rate more credible.
The bias current $I$ (see Fig.~\ref{figCircuit}) is chosen from these considerations. 
The ratio
\begin{equation}
	\beta = I/I_0
\end{equation}
is given in Table~\ref{tablParameters}.

One needs to know the position of the levels in the well.
For that, we expand the potential around the well minimum up to the cubic terms:
\begin{equation}
	\frac W \Wj \approx \frac{\sqrt{1-\beta^2}}2 \delta^2 - \frac{\beta}6 \delta^3,
\end{equation}
where $\delta = 2\pi\Phi/\Phi_0 - \phi_\min$ is the dimensionless flux with respect to the well minimum at $\phi_\min = \arcsin\beta$.
To determine the level structure correctly, the cubic approximation should be accurate in the region up to the barrier maximum at $\delta_\max = 2\cot\phi_\min$.
For a weak anharmonicity, one can calculate the position of the levels using the second order perturbation theory~\cite{landau1991quantum,strauch2004theory}.
It is useful to define
\begin{equation}
\label{eqn0}
	n_0 = \frac{(1-\beta^2)^{5/4}}{3\beta^2} \sqrt{\frac\Wj{2\Wc}}
\end{equation}
the barrier height in the units of
\begin{equation}
\label{eqPlasmaFrequency}
	\omega_\plasma = \frac 1 {\hbar} \sqrt{8\Wj\Wc} (1-\beta^2)^{1/4}
\end{equation}
the level separation in harmonic approximation.
Expressions~\eqref{eqn0} and \eqref{eqPlasmaFrequency} coincide with those given in Ref.~\cite{strauch2004theory}.
Transition frequency from the ground to the first excited state is $\omega_{10} = \omega_\plasma(1 - 5/36n_0)$.
Transition frequency to the second excited state is $\omega_{20} = \omega_\plasma(2 - 5/12n_0)$~\cite{martinis2003decoherence}.
We aim to detect photons of frequency
\begin{equation}
	\omega = \omega_{02}/2.
\end{equation}
This photon is detuned from the $0\to1$ transition by
\begin{equation}
\label{eqAbsoluteUnharmonicity}
	\Delta = \omega_{10} - \omega
		= \frac {5}{72} \frac {\omega_\plasma} {n_0}.
\end{equation}
We provide the value of $\Delta$ in Table~\ref{tablParameters}.

Knowledge of $\Delta$ allows one to set couplings $g_1$ and $g_2$.
One can use the criterion~\eqref{eqValidityOfEffH} for that.
To be sure that no clicks are delivered when there is a single photon in the resonator, Eq.~\eqref{eqValidityOfEffH} should hold for $\nch=1$.
This requirement does not matter for bigger photon numbers---by the reasoning similar to that in the end of Section~\ref{secError}.
So, we choose $\lambda_2 = 0.1$ which fulfills one of the requirements~\eqref{eqValidityOfEffH} for $\nch = 1$.
By virtue of Eq.~\eqref{eqEstimateRatiog2g1}, the part of the condition~\eqref{eqValidityOfEffH} with $\lambda_1$~\eqref{eqLambda12} holds automatically.
From the definition~\eqref{eqLambda12} of $\lambda_1$ and $\lambda_2$ and the relationship~\eqref{eqEstimateRatiog2g1}, one gets that
\begin{equation}
\label{eqg1g2}
	g_2 = \lambda_2 \Delta,
\quad
	g_1 = \lambda_2 \Delta / \sqrt 2.
\end{equation}

One also needs to make sure that the condition~\eqref{eqSmallStarkShift} holds. This yields the biggest photon number $N_{\text{max}}$ that can be distinguished from the single-photon and the vacuum states.
Its value is given in Table~\ref{tablParameters}.

Let us calculate the rate $B_{20}$ of the two-photon absorption.
It follows from the Eqs.~\eqref{eq20Absorption}, \eqref{eq20Coupling}, \eqref{eqg1g2}, and~\eqref{eqLambda12} that
\begin{equation}
	B_{20}
		\approx 4 \lambda_2^4 \Delta^2 / (\tilde\Gamma_2 + \Gamma_{22}).
\end{equation}
Assuming flat density of states of the thermal reservoir, one can estimate in the harmonic approximation that
\begin{equation}
\label{eqGamma21Gamma10}
	\Gamma_{21} \approx 2\Gamma_{10}.
\end{equation}
Tunneling rates $\gamma_0$, $\gamma_1$, and $\gamma_2$ are calculated with the WKB method~\cite{landau1991quantum} and are given in Table~\ref{tablParameters}.
The action integral was carried out numerically for the exact potential $W$ given by Eq.~\eqref{eqHjpm}.
With all the necessary quantities obtained, one can calculate $B_{20}$.
Its value is provided in the table.

\begin{figure}
\includegraphics{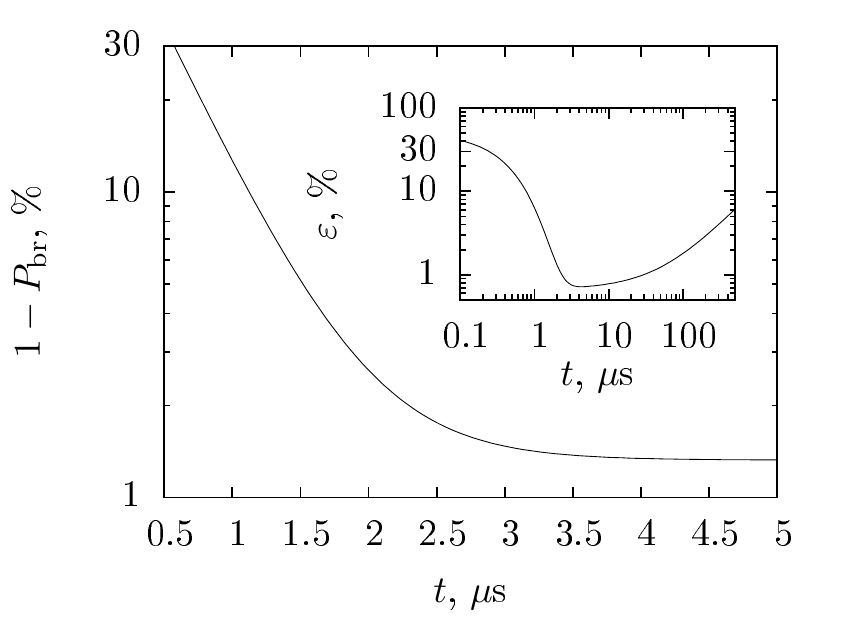}
\caption{Probability of missing a two-photon state for a device with the parameters given in Table~\ref{tablParameters}.
(Inset) Error in the discrimination of the two-photon state against the states with less photons.
}
\label{figError}
\end{figure}

Now one can estimate the JPM performance.
The error~\eqref{eqError} and the probability to miss a two-photon state $1-\Pbright$~[see Eq.~\eqref{eqP}] are shown in Fig.~\ref{figError}.
The dominant contribution to the false counts at the optimal counting time~\eqref{eqOptimalTime} is due to the ground level tunneling as given by Eqs.~\eqref{eqP0} and~\eqref{eqP1}.
The false count probability is given in Table~\ref{tablParameters}.
Transitions to the first excited state $\ket 1$ in the second-order perturbation theory in $\lambda_{1,2}$ contribute as well.
However, one can check that by the criterion~\eqref{eqValidityTimeFirstOrderPerturbation} their effect is still vanishing for the relevant times.
Also, $\ket 1$ could be excited by an off-resonant single photon due the level widening. 
However, this is highly improbable, as
\begin{equation}
\label{eqFirstLevelNotTooFat}
	\tilde\Gamma_1 + \Gamma_{11} \ll \Delta.
\end{equation}

\section{Counting to two}
\label{secCounting}

A two step procedure~(see Fig.~\ref{figHowItWorks} and Section~\ref{secIntro}) is to be performed to count photons to two.
To switch from the two-photon mode to the single-photon one, bias current $I$~[see Fig.~\ref{figCircuit} and the Hamiltonian~\eqref{eqHjpm}] is changed so that the JPM possesses two metastable states instead of three.
Here we estimate the error in discrimination between the vacuum input state, a one-photon state, and a multi-photon state.
The total time of the discrimination is estimated as well.

Full time to count to two is approximately the same as the time to distinguish a multi-photon state vs. the vacuum or the single-photon state.
Additional time consists of the time to switch to the single-photon mode and the time to discriminate the vacuum state.
To spur no excitations in the JPM, the switching should be much slower then the inverse transition frequencies.
For the parameters in Table~\ref{tablParameters}, the switching can be as fast as \SI{10}{\ns}.
Now let us compare the waiting times.
Time to discriminate a multi-photon state is determined by $B_{20}$, as it follows from Eq.~\eqref{eqP}.
Time to discriminate the vacuum is set by $\gamma_1$ in the two-state configuration and $B_{10}$ the single photon absorption rate.
The latter can be calculated analogously to the two-photon absorption rate~\eqref{eq20Absorption}.
This yields $B_{10} = 4 g_1^2 / (\tilde\Gamma_1 + \Gamma_{11})$.
With Eq.~\eqref{eq20Coupling} one has
\begin{equation}
\label{eq20Absorption10Absorption}
	\frac{B_{20}}{B_{10}}
		= 2\lambda_2^2 \frac{\tilde\Gamma_1 + \Gamma_{11}}
							{\tilde\Gamma_2 + \Gamma_{22}}
		\approx \lambda_2^2.
\end{equation}
Equation~\eqref{eqGamma21Gamma10} and the fact that decay is much faster than the pure decoherence were used in obtaining the last equality in Eq.~\eqref{eq20Absorption10Absorption}.
WKB estimate for the tunneling rate from the excited state gives
\begin{equation}
\label{eqg1}
	\gamma_1 \approx 2\pi\,\SI{19}{\MHz}.
\end{equation}
We chose $I/I_0 = 0.98473$ to fit two levels in the well.
For this choice, the excited level is not situated very close to the top of the barrier.
By Eq.~\eqref{eqg1} and the value for $B_{20}$ from Table~\ref{tablParameters}, as well as from Eq.~\eqref{eq20Absorption10Absorption}, discrimination of the vacuum state is much faster than that of a multi-photon state.

Now we find the probability to incorrectly determine the number of input photons.
Let $\Pbright^{0/1}$ denote the probability to correctly identify a single-photon state in the second stage;
$\Pbright$ denotes that in the first stage as before.
Probability of error in the two-step discrimination is then
\begin{multline}
\label{eqErrorCountingRaw}
	\fullerror = P_0 \Pfalse
				+ P_1 [\Pfalse + (1-\Pfalse)(1-\Pbright^{0/1})]
				+ P_2 (1 - \Pbright).
\end{multline}
It was taken into account that the false count probability is negligible for the second stage, as compared to $\Pfalse$ the false count probability in first stage.
This is due to the detection time in the second stage being much smaller than that in the first one.
We assume that nothing is known about the input and $P_0 = P_1 = P_2 = 1/3$.
One can rewrite Eq.~\eqref{eqErrorCountingRaw} in a more convenient form:
\begin{equation}
\label{eqFullError}
	\fullerror = \tfrac 1 3 (\Pfalse(1+\Pbright^{0/1}) 
				 + 1-\Pbright^{0/1} + 1-\Pbright).
\end{equation}
To compute $\fullerror$, one needs to estimate $\Pbright^{0/1}$.
For the optimal counting time, a photon is most probably absorbed by the JPM.
However, this does not necessarily gives a click:
a photon could get stuck in the ground state due to the JPM relaxation with a probability $\Gamma_{10} / (\Gamma_{10} + \gamma_1)$.
Therefore,
\begin{equation}
\label{eqPbr01}
	\Pbright^{0/1} \approx \gamma_1 / (\Gamma_{10} + \gamma_1) \approx 98.3\%,
\end{equation}
where the estimate~\eqref{eqg1} was used.
The expression~\eqref{eqPbr01} was also given in Ref.~\cite{poudel2012efficiency}.
With the estimate~\eqref{eqPbr01} and the values from Table~\ref{tablParameters} one obtains
\begin{equation}
	\fullerror \approx 1.1\%.
\end{equation}
The optimal time could be chosen to minimize the full counting error~\eqref{eqFullError} instead of that in the discrimination of a multi-photon state~\eqref{eqError}.
However, this does not improve the full error substantially.

\section{Discussion}

We have proposed a detector of microwave photons with limited photon number resolution.
Realistic parameters have been provided that enable decent performance of the device.
We have evaluated the probability of an error in counting photons and the time needed for the measurement.
The most time consuming part in the device operation is the discrimination of a multi-photon state vs. the single-photon or vacuum one.
The speed of this step is limited by the two-photon absorption rate, which in turn is set by the coupling strength of the JPM to the cavity.
To avoid single-photon transitions, the coupling should be much weaker than the JPM anharmonicity.
A larger anharmonicity can lead to faster detection.
Moreover, faster detection decreases the false count probability $\Pfalse$.
The probability to count photons incorrectly is determined by $\Pfalse$ and the probabilities to miss a multi- and the single-photon state.
As for the probabilities to miss photons, they are determined by branching ratios between the excited state tunneling and relaxation.

For the proposed parameters, the \SI{8.2}{\GHz} photons are detected.
The frequency could be chosen at the design stage in the range from \SI 1 {\GHz} to \SI{20}{\GHz}.
The upper limit on the frequency is set by the superconducting gap of aluminum, which is about~\SI{82}{\GHz}, and the condition~\eqref{eqBigGap}.
As for the lower limit, it is determined by the requirement~\eqref{eqFirstLevelNotTooFat}, the relationship~\eqref{eqAbsoluteUnharmonicity} between the plasma frequency and the anharmonicity, and an estimate for the decoherence of the JPM first excited state, which is about~\SI 1 {\MHz}.

Two possibilities for development of the detector are worth mentioning.
First, one can use it for detection of itinerant photons.
One option is to attach the resonator to a waveguide;
it will function as a capture cavity from Ref.~\cite{opremcak2018measurement}.
Another option is to attach a waveguide directly.
This introduces reflection losses; to minimize them, the detector should be matched to its input~\cite{schoendorf2018optimizing}.
Moreover, we need to perform two stages: the detection of a multi-photon state and the detection of vacuum.
This can be accomplished by two devices connected to a waveguide in sequence.
Secondly, one can envision a detector that counts photons up to $N$.
This device might use an $N$-photon transition through $N-2$ auxiliary levels at the first stage to discriminate the states with $N$ or more photons.
Afterwards, it can be sequentially tuned to discriminate the states with $N-1$, $N-2$, and down to 1 photon.

\begin{acknowledgments}
The work of A.S.~was supported by DAAD scholarship (2016).
We thank Robert McDermott for useful comments on the manuscript.
\end{acknowledgments}

\appendix

\section{Derivation of the circuit Hamiltonian}
\label{apHamiltonian}

Lagrangian of the system is given by
\begin{gather}
	L = L_\jpm + L_\res + L_\coup,
\\
	L_\jpm = \frac{C\dot\Phi^2}2 + \Wj\cos 2\pi\frac\Phi{\Phi_0} + I\Phi,
\\
	L_\coup = \frac{C'(\dot\Phi - \dot\Phi_\res)^2}2,
\\
	L_\res = \frac{C_\res\dot\Phi_\res^2}2 - \frac{\Phi_\res^2}{2L_\res}.
\end{gather}
Here $\Wj$ is defined by Eq.~\eqref{eqWj}.

Generalized momenta are
\begin{gather}
\label{eqGeneralizedMomentum}
	Q = \partial L / \partial\dot\Phi = (C + C')\dot\Phi - C'\dot\Phi_\res,
\\
	Q_\res = \partial L / \partial\dot\Phi_\res 
		   = -C'\dot\Phi + (C_\res + C')\dot\Phi_\res.
\end{gather}

The system Hamiltonian is given by the Legendre transform,
\begin{equation}
\label{eqLegendre}
	H = Q \dot\Phi + Q_\res \dot\Phi_\res - L.
\end{equation}
One needs to find the kinetic energy part $T$ of $H$.
It is a quadratic form in $Q$ and $Q_\res$,
\begin{equation}
\label{eqT}
	T = \frac 1 2 \frac{\partial^2 H}{\partial Q^2} Q^2
		+ \frac 1 2 \frac{\partial^2 H}{\partial Q_\res^2} Q_\res^2
		+ \frac{\partial^2 H}{\partial Q \partial Q_\res} Q Q_\res.
\end{equation}
It was used that potential energy, which composes the rest of $H$, does not depend on the momenta.
Differentiating Eq.~\eqref{eqLegendre} and using the definition of generalized momentum~\eqref{eqGeneralizedMomentum} gives rise to
\begin{equation}
	\partial^2 H / \partial Q^2 = \partial \dot\Phi / \partial Q.
\end{equation}
Other coefficients are given by the similar formulas.
One then determines the renormalized capacitances
\begin{gather}
	\tilde C = \frac{C + C'(1 + C/C_\res)}{1 + C'/C_\res},
\quad
	\tilde C_\res = \frac{C_\res + C'(1 + C_\res/C)}{1 + C'/C},
\\
	\tilde C' = (1/C' + 1/C + 1/C_\res)^{-1}
\end{gather}
in the Hamiltonian~\eqref{eqHamiltonian}--\eqref{eqCouplingHamiltonian}.

\section{Dressed Lindblad equation}
\label{apLindblad}

Here we write out explicit form of the first-order corrections in the Lindbladian in our working frame.

On applying the transform $U$~\eqref{eqSchriefferWolffTransform}, the Lindbladian~\eqref{eqLindbladBare} becomes
\begin{equation}
\label{eqLindbladianDressed}
	L \to L + L^{(1)},
\end{equation}
where $L^{(1)}$ is first-order in $\lambda_1$ and $\lambda_2$.
It can be given in terms of its matrix elements:
\begin{widetext}
\input{Lcorij}
\end{widetext}

\bibliography{common_sources,jj_sources,cqed,counters,comp,amp,discrimination,array,few_photon_sources}
\bibliographystyle{apsrev4-1}

\end{document}

%% file: fizychni_komandy.tex
\usepackage{amsmath}

\newcommand{\hc}{\text{H.~c.}}
\newcommand{\cc}{\text{c.~c.}}

\newcommand{\bra}[1]{\langle{#1}|} 
\newcommand{\ket}[1]{|{#1}\rangle} 
\newcommand{\Bra}[1]{\left\langle{#1}\right|} 
\newcommand{\Ket}[1]{\left|{#1}\right\rangle} 



%% file: notations.tex
\newcommand{\jpm}{\mathrm{JPM}}
\newcommand{\res}{\mathrm{r}}
\newcommand{\coup}{\mathrm{c}}

\newcommand{\Wc}{E_\mathrm{C}}
\newcommand{\Wj}{E_\mathrm{J}}

\newcommand{\plasma}{\mathrm{p}}

\newcommand{\me}{\mathrm m}

\newcommand{\bare}{\mathrm b}
\newcommand{\inter}{\mathrm i}

\newcommand{\sigmaz}[1]{\sigma_\mathrm{z}^{#1}}
\newcommand{\nch}{N_\mathrm{ch}}

\newcommand{\Pfalse}{P_{\mathrm f}}
\newcommand{\Pbright}{P_{\mathrm b}}

\newcommand{\fullerror}{\varepsilon^{0/1/2}}

%% file: Lcorij.tex
\begin{dgroup*} 
\begin{dmath}\Bra{0} L^{(1)} \Ket{0}=\lambda_2\,\Gamma_{10}\,a^\dag 
 \Bra{2} \rho \Ket{1}-{{1}\over{2}}\,\lambda_1\,\left(\gamma_1-
 \gamma_0+\Gamma_{11}+\Gamma_{10}\right)\,a^\dag \Bra{1} \rho \Ket{0}
 +\lambda_1\,\Gamma_{10}\,a \Bra{0} \rho \Ket{1}+\cc,\end{dmath}
\begin{dmath}\Bra{1} L^{(1)} \Ket{1}={{1}\over{2}}\,\lambda_2\,\left(
 \gamma_2-\gamma_1+\Gamma_{22}+\Gamma_{21}+\Gamma_{11}-\Gamma_{10}
 \right)\,a^\dag \Bra{2} \rho \Ket{1}-\lambda_2\,\Gamma_{21}\,a 
 \Bra{1} \rho \Ket{2}+{{1}\over{2}}\,\lambda_1\,\left(-\gamma_1+
 \gamma_0+\Gamma_{11}-\Gamma_{10}\right)\,a \Bra{0} \rho \Ket{1}+\cc,\end{dmath}
\begin{dmath}\Bra{2} L^{(1)} \Ket{2}={{1}\over{2}}\,\lambda_2\,\left(
 \gamma_2-\gamma_1-\Gamma_{22}+\Gamma_{21}-\Gamma_{11}-\Gamma_{10}
 \right)\,a \Bra{1} \rho \Ket{2}+\cc,\end{dmath}
\begin{dmath}\Bra{\mathrm m} L^{(1)} \Ket{\mathrm m}=\lambda_2\,
 \gamma_1\,a^\dag \Bra{2} \rho \Ket{1}-\lambda_1\,\gamma_0\,a^\dag 
 \Bra{1} \rho \Ket{0}-\lambda_2\,\gamma_2\,a \Bra{1} \rho \Ket{2}+
 \lambda_1\,\gamma_1\,a \Bra{0} \rho \Ket{1}+\cc,\end{dmath}
\begin{dmath}\Bra{0} L^{(1)} \Ket{1}=\lambda_1\,\Gamma_{21}\,a^\dag 
 \Bra{2} \rho \Ket{2}-{{1}\over{2}}\,\lambda_1\,\left(\gamma_1-
 \gamma_0-\Gamma_{11}+\Gamma_{10}\right)\,a^\dag \Bra{1} \rho \Ket{1}
 -\lambda_1\,\Gamma_{10}\,\Bra{1} \rho a^\dag \Ket{1}+{{1}\over{2}}\,
 \lambda_1\,\left(-\gamma_1+\gamma_0-\Gamma_{11}-\Gamma_{10}\right)\,
 \Bra{0} \rho a^\dag \Ket{0}+{{1}\over{2}}\,\lambda_2\,\left(\gamma_2
 -\gamma_1+\Gamma_{22}+\Gamma_{21}-\Gamma_{11}-\Gamma_{10}\right)\,
 \Bra{0} \rho a \Ket{2},\end{dmath}
\begin{dmath}\Bra{1} L^{(1)} \Ket{2}={{1}\over{2}}\,\lambda_2\,\left(
 \gamma_2-\gamma_1-\Gamma_{22}+\Gamma_{21}-\Gamma_{11}-\Gamma_{10}
 \right)\,a^\dag \Bra{2} \rho \Ket{2}-{{1}\over{2}}\,\lambda_1\,
 \left(\gamma_1-\gamma_0+\Gamma_{11}+\Gamma_{10}\right)\,a \Bra{0} 
 \rho \Ket{2}+\lambda_2\,\Gamma_{21}\,\Bra{2} \rho a^\dag \Ket{2}+{{1
 }\over{2}}\,\lambda_2\,\left(\gamma_2-\gamma_1+\Gamma_{22}+
 \Gamma_{21}+\Gamma_{11}-\Gamma_{10}\right)\,\Bra{1} \rho a^\dag 
 \Ket{1},\end{dmath}
\begin{dmath}\Bra{0} L^{(1)} \Ket{2}={{1}\over{2}}\,\lambda_2\,\left(
 \gamma_2-\gamma_1+\Gamma_{22}+\Gamma_{21}-\Gamma_{11}-\Gamma_{10}
 \right)\,\Bra{0} \rho a^\dag \Ket{1}-{{1}\over{2}}\,\lambda_1\,
 \left(\gamma_1-\gamma_0+\Gamma_{11}+\Gamma_{10}\right)\,a^\dag 
 \Bra{1} \rho \Ket{2},\end{dmath}
\begin{dmath}\Bra{\mathrm m} L^{(1)} \Ket{0}={{1}\over{2}}\,\lambda_1
 \,\left(-\gamma_1+\gamma_0-\Gamma_{11}-\Gamma_{10}\right)\,
 \Bra{\mathrm m} \rho a \Ket{1},\end{dmath}
\begin{dmath}\Bra{\mathrm m} L^{(1)} \Ket{1}={{1}\over{2}}\,\lambda_2
 \,\left(\gamma_2-\gamma_1+\Gamma_{22}+\Gamma_{21}-\Gamma_{11}-
 \Gamma_{10}\right)\,\Bra{\mathrm m} \rho a \Ket{2}-{{1}\over{2}}\,
 \lambda_1\,\left(\gamma_1-\gamma_0+\Gamma_{11}+\Gamma_{10}\right)\,
 \Bra{\mathrm m} \rho a^\dag \Ket{0},\end{dmath}
\begin{dmath}\Bra{\mathrm m} L^{(1)} \Ket{2}={{1}\over{2}}\,\lambda_2
 \,\left(\gamma_2-\gamma_1+\Gamma_{22}+\Gamma_{21}-\Gamma_{11}-
 \Gamma_{10}\right)\,\Bra{\mathrm m} \rho a^\dag \Ket{1}.\end{dmath}
\end{dgroup*} 